\documentclass[a4paper,11pt]{article}

\usepackage{jinstpub} 

\usepackage{lineno}

\usepackage{subcaption}
\usepackage{graphicx}
\usepackage{array, multirow}

\title{\boldmath Design, fabrication and large scale qualification of cosmic muon veto scintillator detectors}

\author[a,1]{Mandar Saraf,\note{Corresponding author.}}
\author[a]{Pandi Raj Chinnappan,}
\author[a]{Aditya Deodhar,}
\author[a,b]{Mamta Jangra,}
\author[a]{J. Krishnamoorthi,}
\author[a]{Gobinda Majumder,}
\author[a]{Veera Padmavathy,}
\author[a]{K.C. Ravindran,}
\author[a, b]{Raj Shah,}
\author[a]{Ravindra Shinde,}
\author[a]{and B. Satyanarayana}

\affiliation[a]{Department of High Energy Physics, Tata Institute of Fundamental Research,\\Dr Hoi Bhabha Road, Mumbai, India}
\affiliation[b]{Homi Bhabha National Institute\\Anushaktinagar Mumbai, India}

\emailAdd{mandar@tifr.res.in}

\abstract
{
The INO collaboration is designing a cosmic muon veto detector (CMVD) to cover the mini-ICAL detector which is operational at the IICHEP transit campus, Madurai in South India. The aim of the CMVD is to study the feasibility of building an experiment to record rare events at a shallow depth of around 100\,m, and use plastic scintillators to veto atmospheric muons from those produced by the rare interactions within the target mass of the detector. The efficiency of such a veto detector should be better than 99.99\% and false positive rate should be less than $10^{-5}$.

The CMVD is being built using extruded plastic scintillator (EPS) strips to detect and tag atmospheric muons. More than 700 EPS strips are required to build the CMVD. Two EPS strips are pasted together to make a di-counter (DC) and wavelength shifting fibres are embedded inside the EPS strips to trap the scintillation light generated by a passing cosmic ray muon and transmit it as secondary photons to the Silicon Photo-Multipliers (SiPMs) mounted at the two ends of the DCs. Since the efficiency requirement of the veto detector is rather high, it is imperative to thoroughly test each and every component used for building the CMVD. A cosmic ray muon telescope has been setup using the DCs to qualify all the DCs that will be fabricated. In this paper we will discuss the details of the design and fabrication of the DCs, the cosmic muon setup and the electronics used for their testing and the test results.
}


\begin{document}
\maketitle
\flushbottom

\section{Introduction}
\label{sec:intro}
The India-based Neutrino Observatory (INO) collaboration has proposed a 50\,kT magnetised Iron Calorimeter (ICAL) detector, to study the mass hierarchy and mixing parameters of atmospheric neutrinos~\cite{ino}. The ICAL will consist of 150 layers of Resistive Plate Chambers (RPCs) as active detector elements, sandwiched between 56\,mm thick iron plates. The ICAL will be placed inside an underground laboratory in a cavern under a mountain near Pottipuram in South India. The tall mountain will provide a rock cover of more than 1.3\,km, corresponding to an overburden of $\sim$3500\,mwe on all the sides of the ICAL, and thus reduce the atmospheric muon background by an order of $10^6$~\cite{muonflux}. This reduction in the muon flux is essential for the detection of neutrinos because neutrinos will be detected indirectly through the muons they form after interaction with the target mass, and only a handful of neutrino interactions may happen in a day. Having the ICAL detector at a shallow depth of say $\sim$100\,m could make many more sites available. In order to detect neutrinos at a shallow depth of $\sim$100\,m, where the cosmic muon flux will be reduced by a factor of only $\sim$$10^2$, a muon veto detector with detection efficiency 99.99\% can efficiently veto cosmic ray muons, so that after the cosmic ray veto only about one in $10^6$ muons will contribute to the background which is similar to that due to the cosmic muon flux in the underground lab at a depth of $\sim$1.3\,km.

The feasibility of constructing a cosmic muon veto detector (CMVD), is being studied by building one around the mini-ICAL detector in the IICHEP laboratory. The mini-ICAL is a scaled-down – by a factor of 1/625 – version of the ICAL detector, operational at the IICHEP lab since mid 2018. It was built to gain experience in the construction of a large-scale electromagnet, to study the detector performance and to test the ICAL electronics in the presence of a fringe magnetic field. This detector is made of a 4\,m $\times$ 4\,m $\times$ 1.1\,m magnet consisting of 11 layers of 56\,mm steel plates with 20 RPCs interleaved between them. Each layer has 2 RPCs arranged in the central region of the magnet. The CMVD will be built using plastic scintillators and it will be imperative to qualify every scintillator strip used to build the CMVD.

In this paper, we describe the testing methodology and the test setup used to qualify the scintillator strips.

\section{Construction of the CMVD} \label{cmvd construction}
\begin{figure}[h]
     \centering
     \begin{subfigure}[b]{0.49\textwidth}
         \centering
         \includegraphics[width=\textwidth]{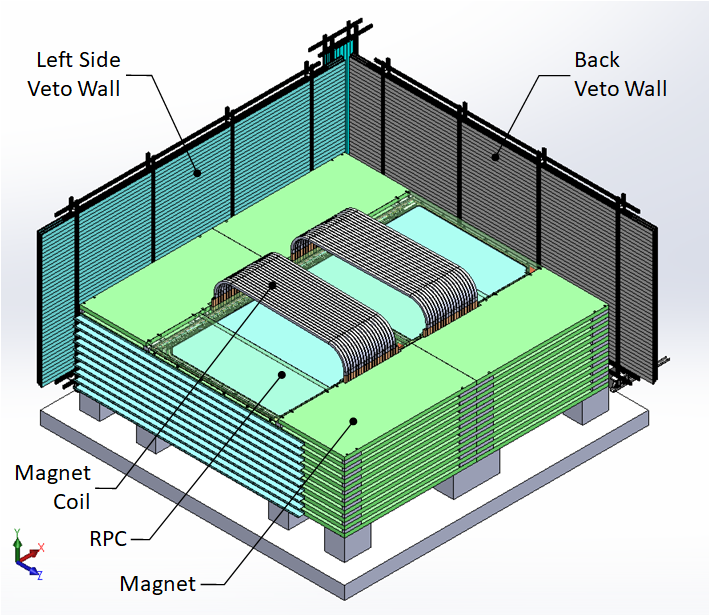}
         \caption{{\footnotesize mini-ICAL with only the left side and back veto walls.}}
         \label{fig:mical:1}
     \end{subfigure}
     \hfill
     \begin{subfigure}[b]{0.49\textwidth}
         \centering
         \includegraphics[width=\textwidth]{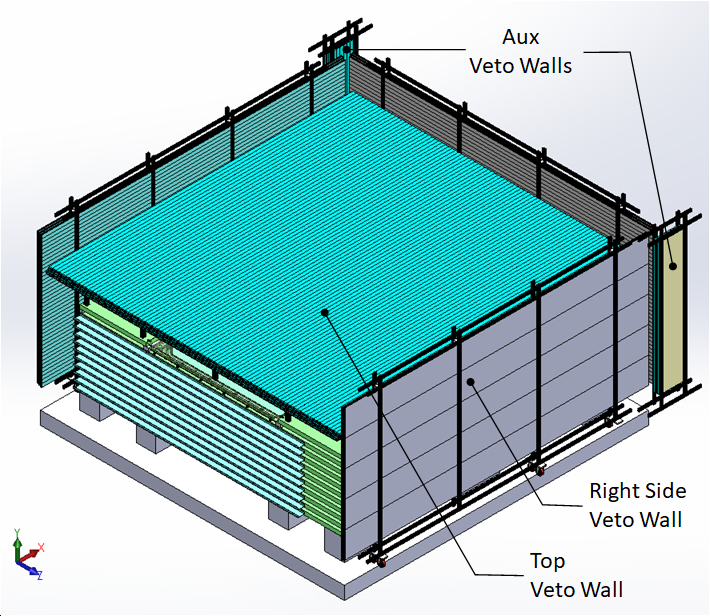}
         \caption{{\footnotesize mini-ICAL with all the veto walls.}}
         \label{fig:mical:2}
     \end{subfigure}
        \caption{The mini-ICAL detector shown partially and fully covered by the veto walls.}
        \label{fig:mical}
\end{figure}

EPS strips~\cite{EPS} will be used to build the veto walls around three sides (left, back and right) and top of the mini-ICAL. The top wall will be made of four layers and the three side walls will be made of three layers of staggered scintillator strips. The staggering will remove any gap through which a muon can pass to the mini-ICAL detector without being detected by the CMVD. In addition to the top and side walls, two more veto walls will be placed at the junction of the back and the side walls to mitigate the gap between them. These are called auxiliary walls. There is no wall in the front side to allow for maintenance of the mini-ICAL detector. As a result, muons passing through a portion of the mini-ICAL detector, from the front, will have to be ignored for the CMVD study. A simulation study is being carried out to determine what portion of mini-ICAL needs to be ignored for the purpose of the CMVD feasibility study. Figure~\ref{fig:mical}(a) shows the mini-ICAL detector with only the left and back veto walls so that the detector is seen clearly and Figure~\ref{fig:mical}(b) shows the mini-ICAL detector fully covered by the veto walls.

For different walls of the CMVD, different lengths of EPS strips are required. This is dictated by the geometry of the mini-ICAL detector being covered. The back wall requires 4.7\,m long, the side walls require 4.6\,m long, the top wall requires 4.5\,m long and the auxiliary walls require 2.1\,m long strips. For the efficiency requirement, 20 mm EPS strips need to be used in the top veto walls. Considering the higher cost of the 20\,mm strips, a balance between cost and efficiency was reached by using a combination of 10 mm and 20 mm thick strips for alternate layers in the top veto wall. The two kinds of strips, the 10\,mm and 20\,mm thick strips, differ slightly in width. The 10\,mm strips are 50\,mm wide, and the 20\,mm strips are 51\,mm wide.

\section{Assembly of the Di-Counters}
\begin{figure}[h]
\centering 
\includegraphics[width=.9\textwidth]{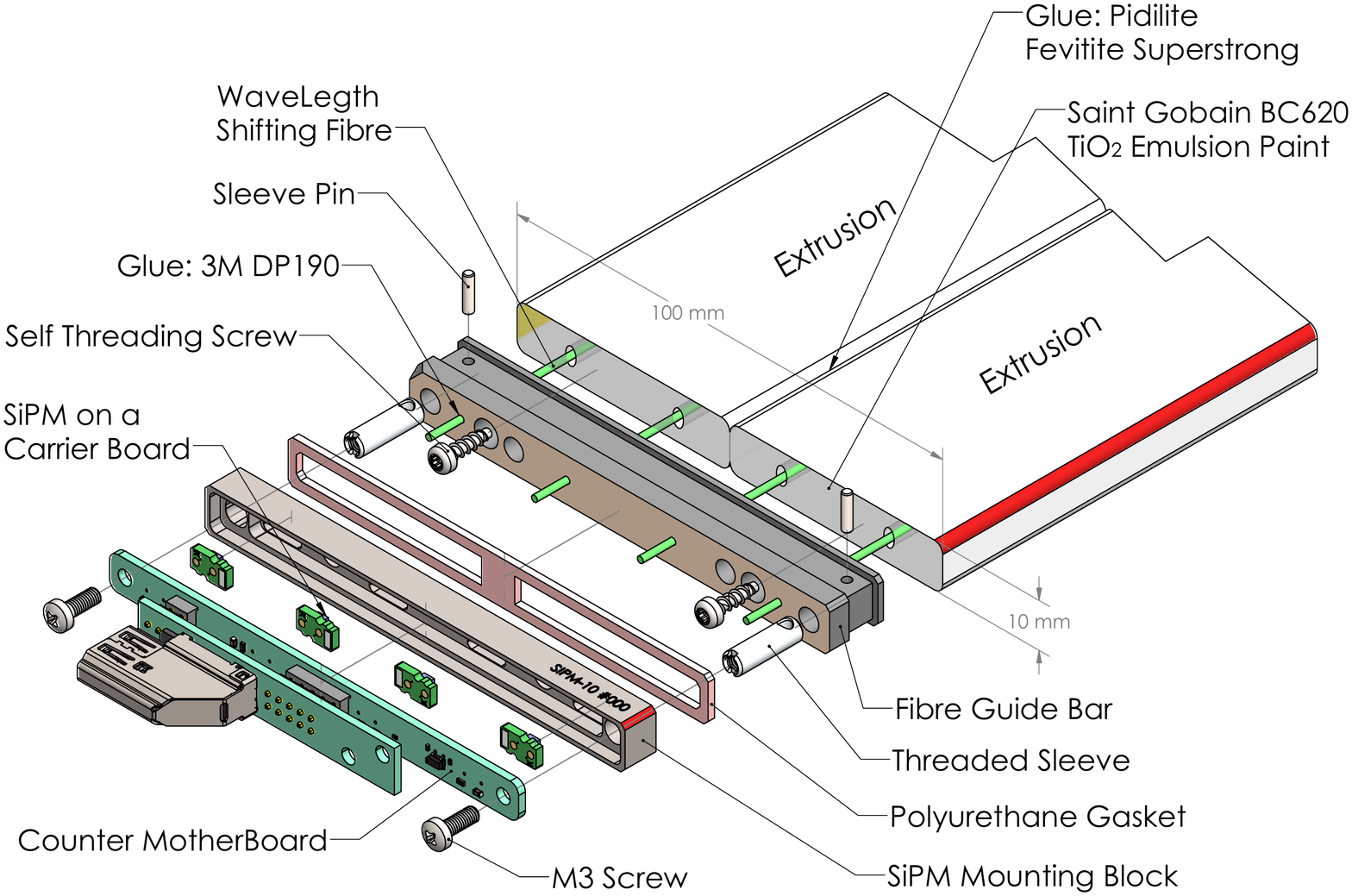}
\caption{\label{fig:dcab} A blown up view of a di-counter assembly.}
\end{figure}

 EPS Strips of different thicknesses and lengths are required for the different veto walls as is described in section \ref{cmvd construction}.  There are 2 holes in the middle of each strip, running along its length, as seen in figure~\ref{fig:dcab}, in which a WaveLength Shifting (WLS) fibre can be inserted. The WLS fibres help in collecting the scintillation photons, produced in the scintillator whenever a charged particle like a muon passes through and interacts with it, and transmit the re-emitted longer wavelength photon to the SiPMs. Kuraray Y-11 WLS fibre is being used for this purpose~\cite{Kuraray}. It is a round multi-clad fibre of 1.4\,mm diameter, with peak absorption wavelength of 430\,nm and peak emission wavelength of 475\,nm. Using a precisely designed arrangement made of a plastic Fibre Guide Bar (FGB) and an aluminium SiPM Mounting Block (SMB), the fibres are guided accurately to the centre of the SiPMs placed in the SMB. A blown up view of this arrangement is shown in the figure~\ref{fig:dcab}. The SiPM – S13360-2050VE, manufactured by Hamamatsu, is being used as a photo detector. The active area of this SiPM is 2\,mm $\times$ 2\,mm~\cite{sipmspecs} and it comprises of 1584 pixels (also called micro-cells). The active area is well suited to interface with the WLS fibre of diameter 1.4\,mm.

Every EPS strip carries two fibres and each fibre is readout by one SiPM on each side. In order to reduce the number of electrical connections from the EPS to the processing electronics, it was decided to combine two EPS strips into a single unit called a di-counter (DC). A DC is constructed by pasting sideways two EPS strips along the length using an epoxy based glue, and then it is cut to the required length. A DC thus has 4 fibres embedded in it and each fibre is readout by two SiPMs, one on each side. At both the ends, fibres are passed through the conical openings in the FGB, glued inside it and the FGB is then attached to the di-counter using self-tapping screws. The surfaces of the fibre and the FGB are made co-planar with the help of a fly-cutter using a diamond tipped tool. An SMB is mounted on each end of the DC using the threaded sleeves and screws. Each SMB houses four SIPMs and a small board (dimensions: 96.6\,mm $\times$ 7.8\,mm), called the Counter Mother Board (CMB) to connect to the SiPMs and transmit the SiPM signals to the processing electronics. The assembly of the di-counter is completed with mounting one CMB on either end.

\begin{table}[h]
\centering
\begin{tabular}{|c | c | c|} 
 \hline
  & Scintillator thickness / Length & Total Di-Counters \\ [0.5ex] 
 \hline\hline
 Back Wall & 10\,mm / 4.7\,m & 60 \\
 \hline
 Right Side Wall & 10\,mm / 4.6\,m & 60 \\
 \hline
 Left Side Wall & 10\,mm / 4.6\,m & 60 \\
 \hline
 \multirow{2}{*}{Top Wall} & 10\,mm / 4.5\,m & 88 \\
 \cline{2-3}
  & 20\,mm / 4.5\,m & 88 \\
 \hline
 Left Aux Wall  & 20\,mm / 2.1\,m & 12 \\
 \hline
 Right Aux Wall & 20\,mm / 2.1\,m & 12 \\
 \hline
  & Total & 380 \\ 
 \hline
\end{tabular}
\caption{Various lengths and thicknesses of di-counters required for the CMVD.}
\label{table:1}
\end{table}

\section{Scintillator Tile Fabrication}
\begin{figure}[h]
    \centering
    \begin{minipage}{.45\textwidth}
        \centering
        \includegraphics[width=0.9\linewidth]{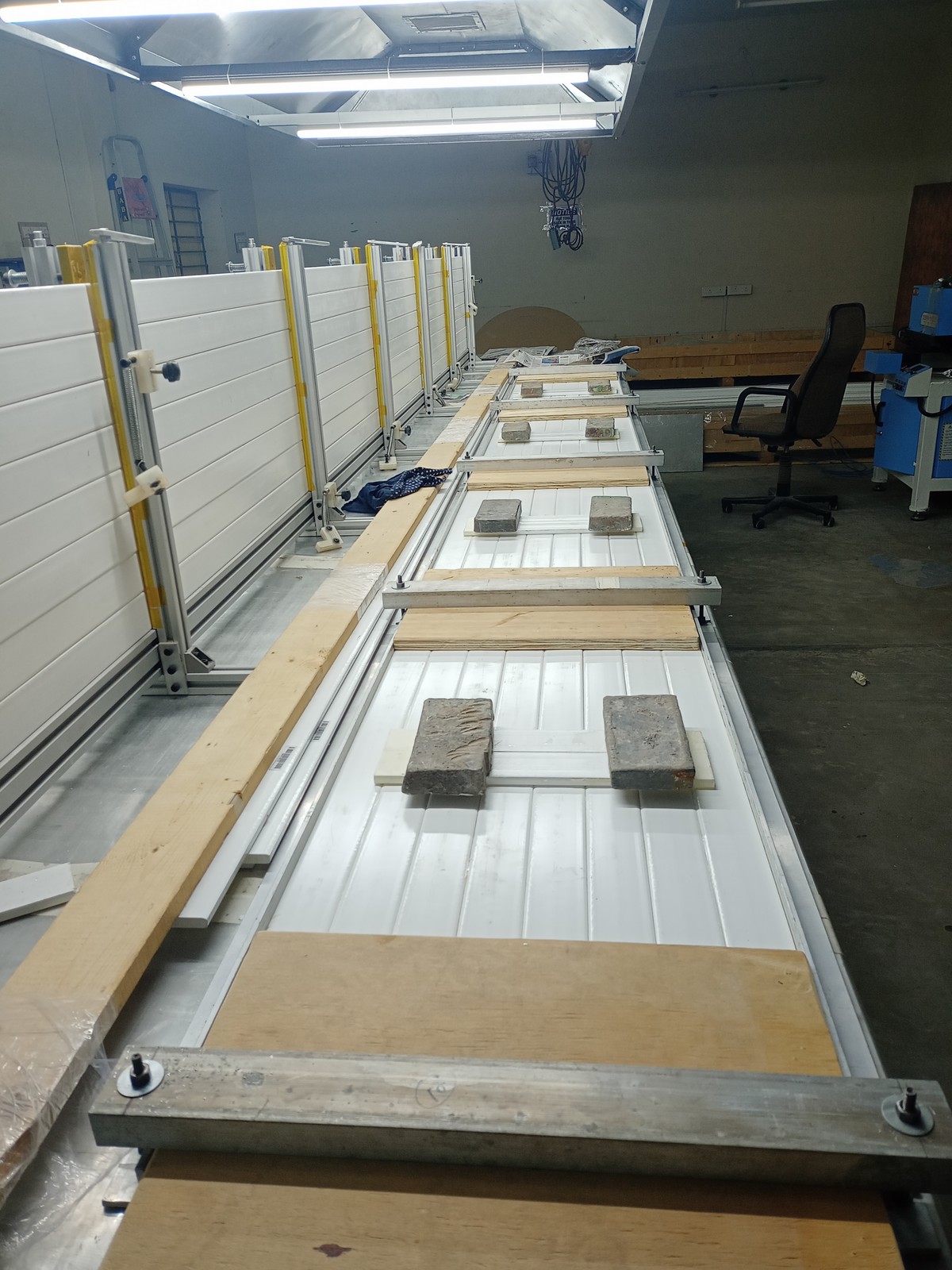}
        \caption{Di-counters glued and clamped to an Al plate for tile fabrication.}
        \label{fig:tile-making-2}
    \end{minipage}
    \hspace{0.05\textwidth}
    \begin{minipage}{0.45\textwidth}
        \centering
        \includegraphics[width=0.9\linewidth]{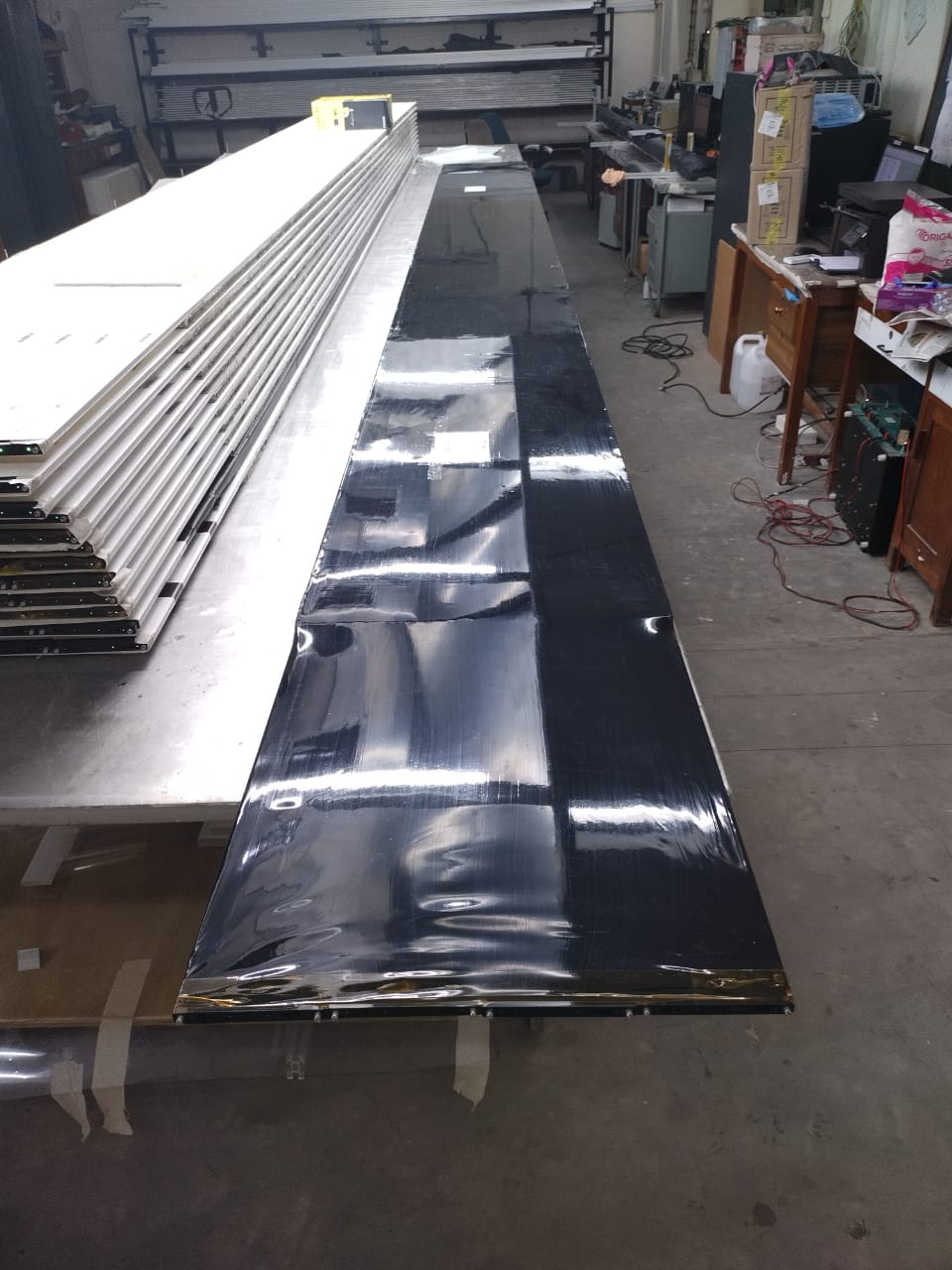}
        \caption{A fully finished tile with the outer PVC sleeve cover.}
        \label{fig:tile-making-3}
    \end{minipage}
\end{figure}

\begin{figure}[h]
\centering 
\includegraphics[width=.75\textwidth]{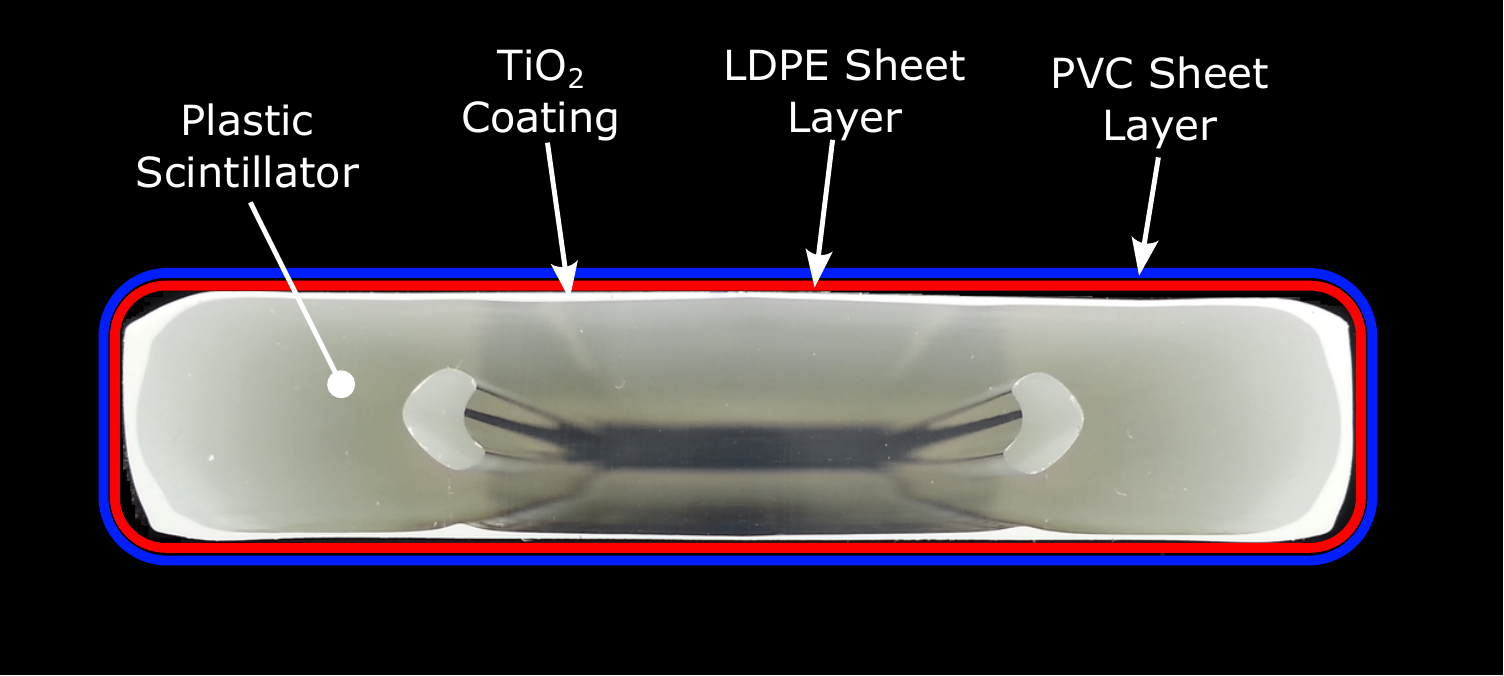}
\caption{\label{fig:dc_wrapping} Schematic of the cross-section of an EPS strip and its protective layers.}
\end{figure}

The DCs are long and thin, and 380 of them will be required to build the veto walls. The various dimensions of di-counters required for the CMVD are enumerated in Table~\ref{table:1}. In order to ease the veto wall construction, it was decided to combine 4 DCs into a single unit called a Tile. A tile is fabricated by gluing 4 DCs onto an Aluminium honeycomb base plate providing mechanical support to the Tile. Figure~\ref{fig:tile-making-2} shows a tile being fabricated. A tile needs to be absolutely light tight in order to meet the criteria of false trigger rate to be less than $10^{-5}$. Since the outer coating of the EPS strips – $TiO_2$ – is not opaque, it allows absorption of light from outside. In order to prevent this light leakage and make the tiles light leak, a sheet of Tedlar paper or black Low-density Polyethylene (LDPE) is wrapped on the tiles. The tiles are further covered with a heat-shrinkable PVC sleeve to prevent any abrasion damage. Figure~\ref{fig:tile-making-3} shows a fully fabricated tile with the outer PVC sleeve covering. Figure~\ref{fig:dc_wrapping} shows the cross-section of an EPS strip, its outer coating of $TiO_2$, and the two external layers of LDPE and PVC respectively.

\section{Di-Counter test setup} \label{dc test setup}
For testing the di-counters, a cosmic ray muon telescope, i.e. a hodoscope, was assembled which also consisted of the di-counters, so that full length coverage could be obtained. The di-counters were tested to evaluate their muon detection efficiency. As mentioned in section~\ref{cmvd construction}, EPS strips of different lengths and thicknesses are going to be used to build the veto walls. Thus in order to have the best geometry, we use the smallest di-counters, i.e. the 10\,mm thick, which are 100\,mm wide and 4.5\,m long for cosmic ray muon trigger generation. 

\begin{figure}[h]
\centering 
\includegraphics[width=.75\textwidth]{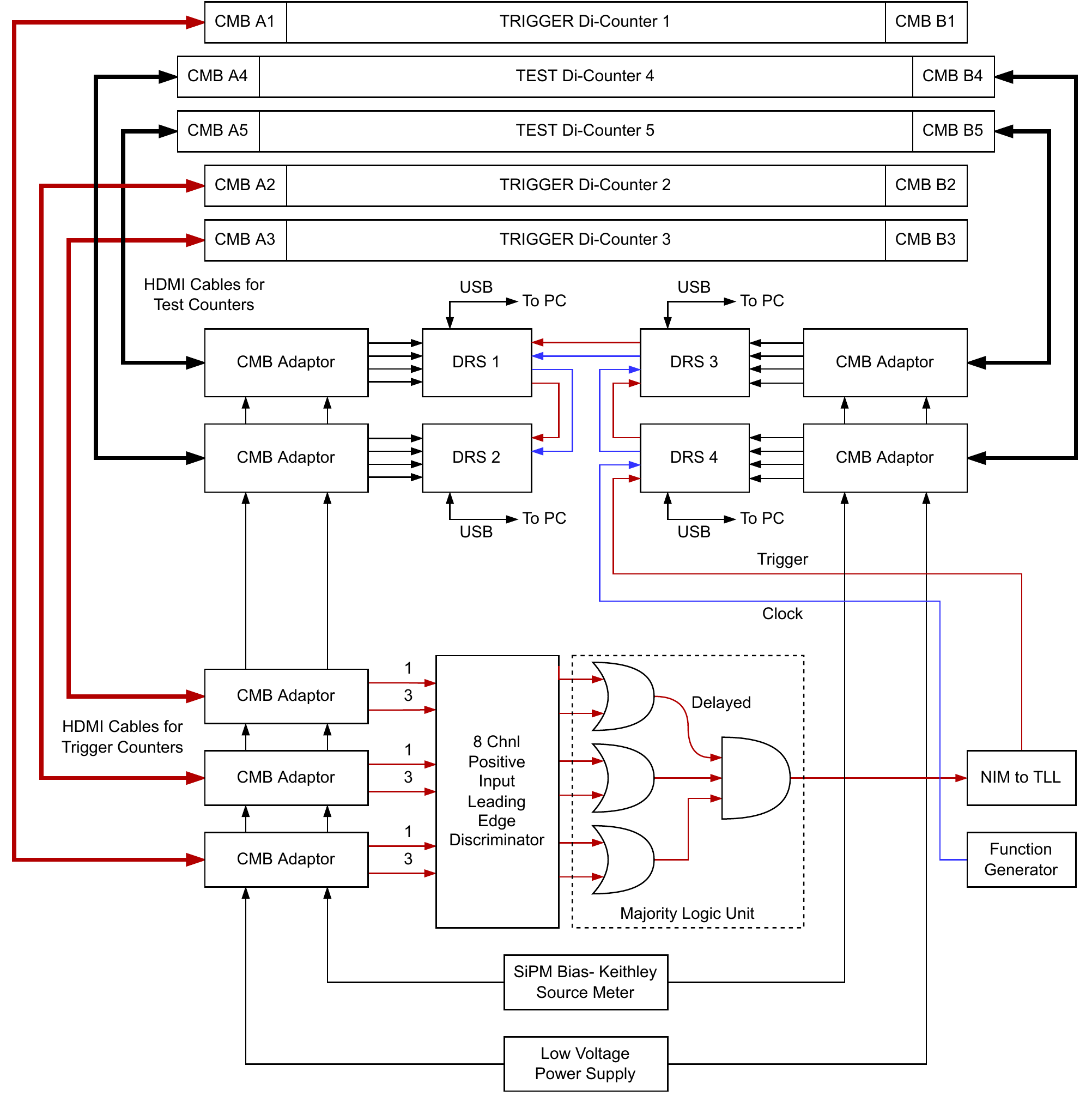}
\caption{\label{fig:dctsc} Di-Counter Qualification Test Setup using Cosmic Ray Muon Telescope.}
\end{figure}

The test setup consists of a stack of di-counters, with two trigger di-counters at the bottom of the stack, and one more trigger di-counter at the top. Two test di-counters are kept in between these. A schematic of this setup is shown in the figure~\ref{fig:dctsc}.

\subsection{Test Setup Electronics}
For signal readout from the SiPMs, a miniature board called the Counter Mother Board (CMB) is used. One CMB is used for signal readout from one side of the di-counter, thus it has to cater to 4 SiPMs. The CMB is located in the SMB, where it interfaces with the SiPM with the help of spring loaded pogo pins. The spring loaded pins also help in maintaining a constant contact between the fibre and the SiPM. In addition to providing bias voltage to the SiPMs and transmitting the SiPM signals to the processing electronics, the CMB board also holds a temperature sensor for precise temperature measurement as well as LEDs and an LED driver circuit for calibration of the SiPMs. The transmission of all the signals, namely, 4 SiPM signals, SiPM bias voltage, LED control signal, temperature sensor readout, and low voltage power supply, between the CMB and the processing electronics, is done using an HDMI cable.

\subsubsection{Amplifier}
The SiPM signals are amplified by a transistor-based trans-impedance amplifier (TIA) followed by an opamp buffer stage to drive the outputs~\cite{mamta2}. The combined gain of this two stage amplifier is ~1245\,$\Omega$. The amplifier is mounted on a board called the CMB-Adaptor board, which also supplies the low voltage and SIPM bias voltage. A Keithley source meter~\cite{keithley} is used to supply the SiPM bias voltage and monitor its current, as it can control the voltage in steps of 5\,mV in the required range of 0-200\,V and measure current with a resolution of 1\,pA (accuracy $\pm$0.3\,nA) in the required range of 0-1\,$\mu$A. For data acquisition, DRS4 evaluation boards~\cite{DRS1, DRS2} have been used.


\subsubsection{DRS - Domino Ring Sampler}
 The Domino Ring Sampler (DRS4) ASIC is a switched capacitor array (SCA) capable of sampling 9 differential input channels at a sampling speed of up to 5 GSPS~\cite{DRS1}. The SCA consists of 1024 capacitive cells per channel to store the input analog waveform. The stored waveform samples can be read out via a shift register clocked at 33 MHz for external digitization. For SIPM data collection we used the DRS4 evaluation board. The evaluation board has four analog inputs, a USB connection for data readout and board configuration and an on-board trigger logic. It is basically equivalent to a four channel 5 GSPS digital oscilloscope. The evaluation boards come with PC software to display as well as store the digital data of the waveforms captured from the four input channels in the PC.

\subsubsection{Trigger Generation}
\label{sec:trigger}
SiPMs are present on both sides of the EPS strips for the readout, but the trigger is generated using SiPMs of only one side, due to the limited number of channels available on the discriminator. One SiPM signal from each EPS strip of the three trigger di-counters, i.e. two SiPM signals from each trigger di-counter, are connected to the inputs of a commercial NIM based discriminator, whose thresholds are set to a voltage equivalent to 3 photo-electrons (p.e.). The discriminator outputs are used to generate a 3-fold coincidence, as shown in figure~\ref{fig:dctsc}, using a commercial NIM based majority logic unit (MLU) to generate a trigger signal which flags a muon event. This trigger signal is then converted to TTL logic to drive the TTL Trigger input of the DRS4 evaluation board.

\subsubsection{Data Acquisition}
Whenever a cosmic ray muon passes through the di-counter test stand, it leaves a signature in the scintillators, eventually causing the SiPMs to generate a pulse. On average, the SiPM pulse for a cosmic ray muon passing through the 10 mm thick scintillator is around 15-20 p.e., and since the threshold for the discriminators is set at 3\,p.e., the pulses pass the discriminator threshold and a trigger signal is formed at the MLU. The trigger signal causes the DRS evaluation boards to acquire and digitize the input signal waveforms of the test di-counters, and these are stored in the computer for further analysis.

\section{Cosmic muon detection efficiency}
\begin{figure}[h]
\centering 
\includegraphics[width=0.99\textwidth]{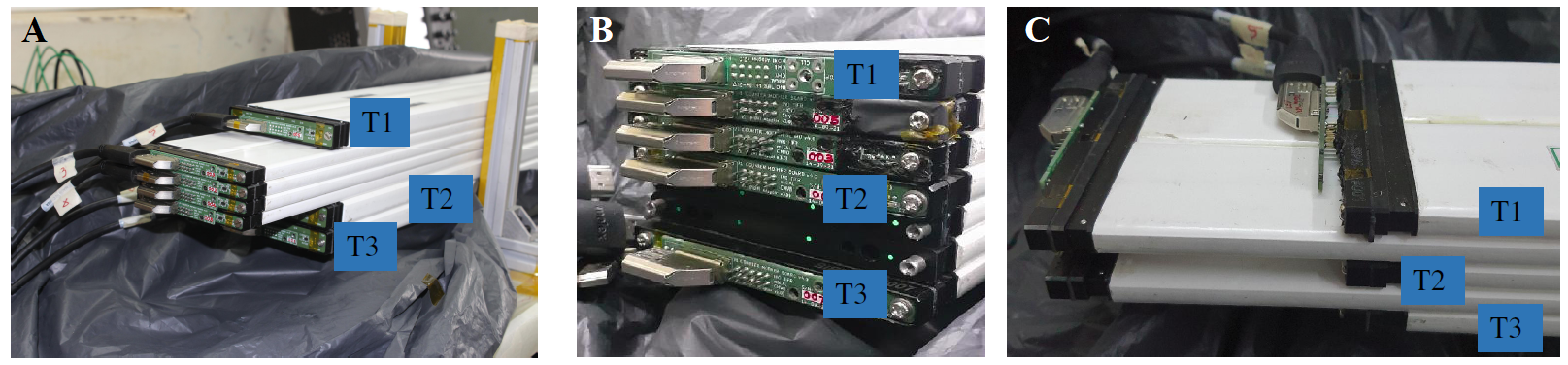}
\caption{\label{fig:geom}Side view of all three geometries from the di-counter testing setup for (a) TYPE1, (b) TYPE2 and (c) TYPE3. Here T1, T2 and T3 are three trigger di-counters.}
\end{figure}

\begin{figure}[h]
\centering 
\includegraphics[width=0.7\textwidth]{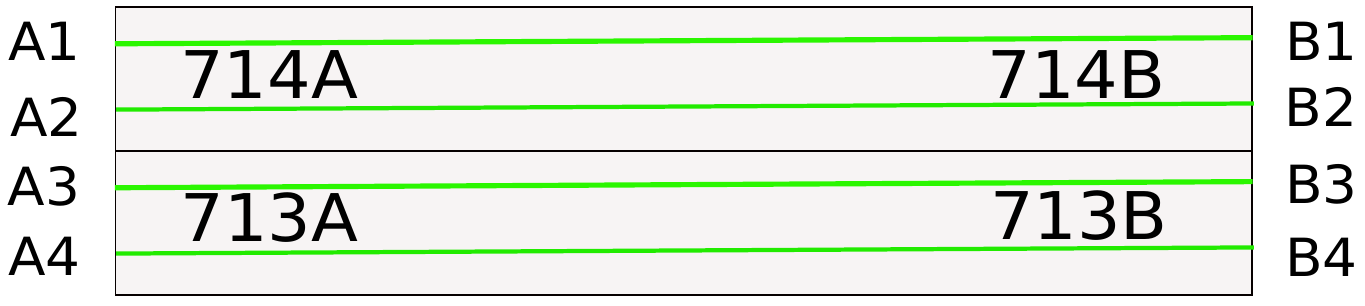}
\caption{\label{fig:dc_sch} Schematic of the di-counter and mounted SiPM sequence.}
\end{figure}

A large number ($\sim$ 400) of di-counters were tested, so different detector geometries were explored to test the di-counters in an effective way. The trigger was formed as mentioned in section~\ref{sec:trigger} such that both the EPS strips could be tested at the same time. The following geometries were explored:
\begin{itemize}
\item \textbf{TYPE1}: As shown in figure~\ref{fig:geom}a, there are two trigger di-counters at the bottom, four test di-counters are in the middle and one more trigger di-counter was placed on the top.
    \begin{itemize}
        \item All the 10 mm di-counters were tested with this geometry.
        \item Trigger Criteria: As is shown schematically in figure~\ref{fig:dc_sch}, the trigger was generated as, $AND$[(A1 $OR$ A3)$_{T1}$, (A1 $OR$ A3)$_{T2}$, (A1 $OR$ A3)$_{T3}$] where T1, T2 and T3 are the three trigger di-counters.
        \item Pairing: For the efficiency calculation from the test di-counters, the SiPM channels were combined as (A1 + A3), (A2 + A4), (B1 + B3) and (B2 + B4).
    \end{itemize}
\item \textbf{TYPE2}: One trigger di-counter is kept at the bottom, this is followed by the first set of two test di-counters, then one more trigger di-counter is placed, then two more test di-counters and finally one more trigger di-counter is placed on the top as shown in figure~\ref{fig:geom}b.
    \begin{itemize}
        \item All the 20 mm di-counters were tested with this geometry.
        \item Trigger Criteria: $AND$[(A1 $OR$ A3)$_{T1}$, (A1 $OR$ A3)$_{T2}$, (A1 $OR$ A3)$_{T3}$].
        \item Pairing: For the efficiency calculation from the test di-counters, the SiPM channels were combined as (A1 + A3), (A2 + A4), (B1 + B3) and (B2 + B4).
    \end{itemize}
\item \textbf{TYPE3}: Two test di-counters are sandwiched between three trigger di-counters as shown in figure~\ref{fig:geom}c.
    \begin{itemize}
        \item Six di-counters were tested with this geometry to validate the random trigger in the TYPE1 geometry.
        \item Trigger Criteria: Trigger was generated as $AND$[(A1 $AND$ A2)$_{T1}$, (A1 $AND$ A2)$_{T2}$, (A1 $AND$ A2)$_{T3}$]. Similarly, trigger was also generated as $AND$( (A3 $AND$ A4)$_{T1}$, (A3 $AND$ A4)$_{T2}$, (A3 $AND$ A4)$_{T3}$ ) and data was taken independently for these two trigger conditions.
        \item Pairing: For the efficiency calculation from the test di-counters; for the first trigger condition above, channels A1, A2, B1 and B2 were considered; for the second trigger condition above, channels A3, A4, B3 and B4 were considered.
    \end{itemize}
\end{itemize}

In the case of the TYPE1 geometry, though four test di-counters were placed simultaneously, the data was taken only for two di-counters at a time because of the limitations on the available DRS boards. A total of $10^{4}$ events were collected for each data set using the cosmic muon coincidence trigger as mentioned in section~\ref{sec:trigger}.
\begin{figure}[h]
\centering 
\includegraphics[width=0.46\textwidth]{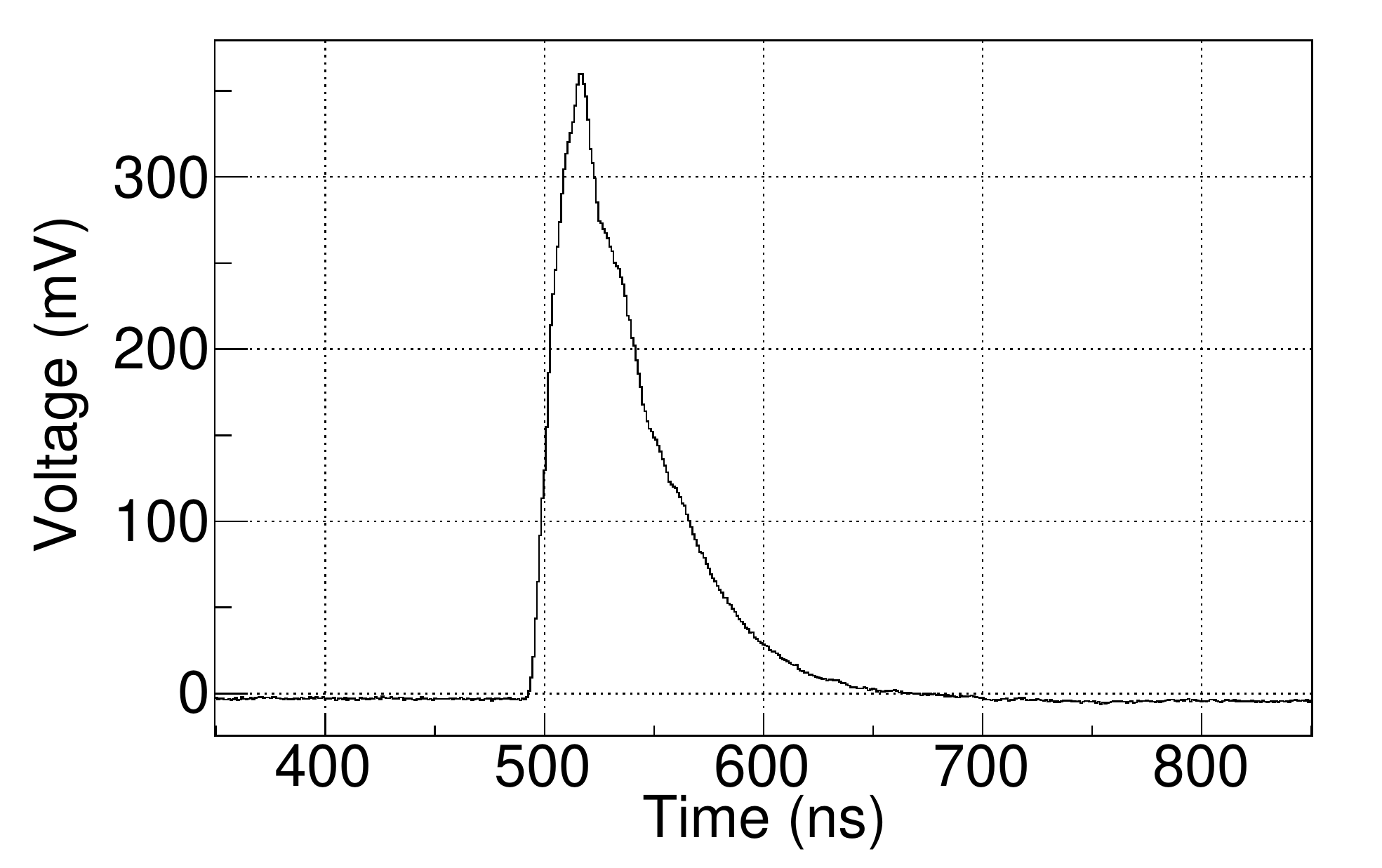}
\includegraphics[width=0.46\textwidth]{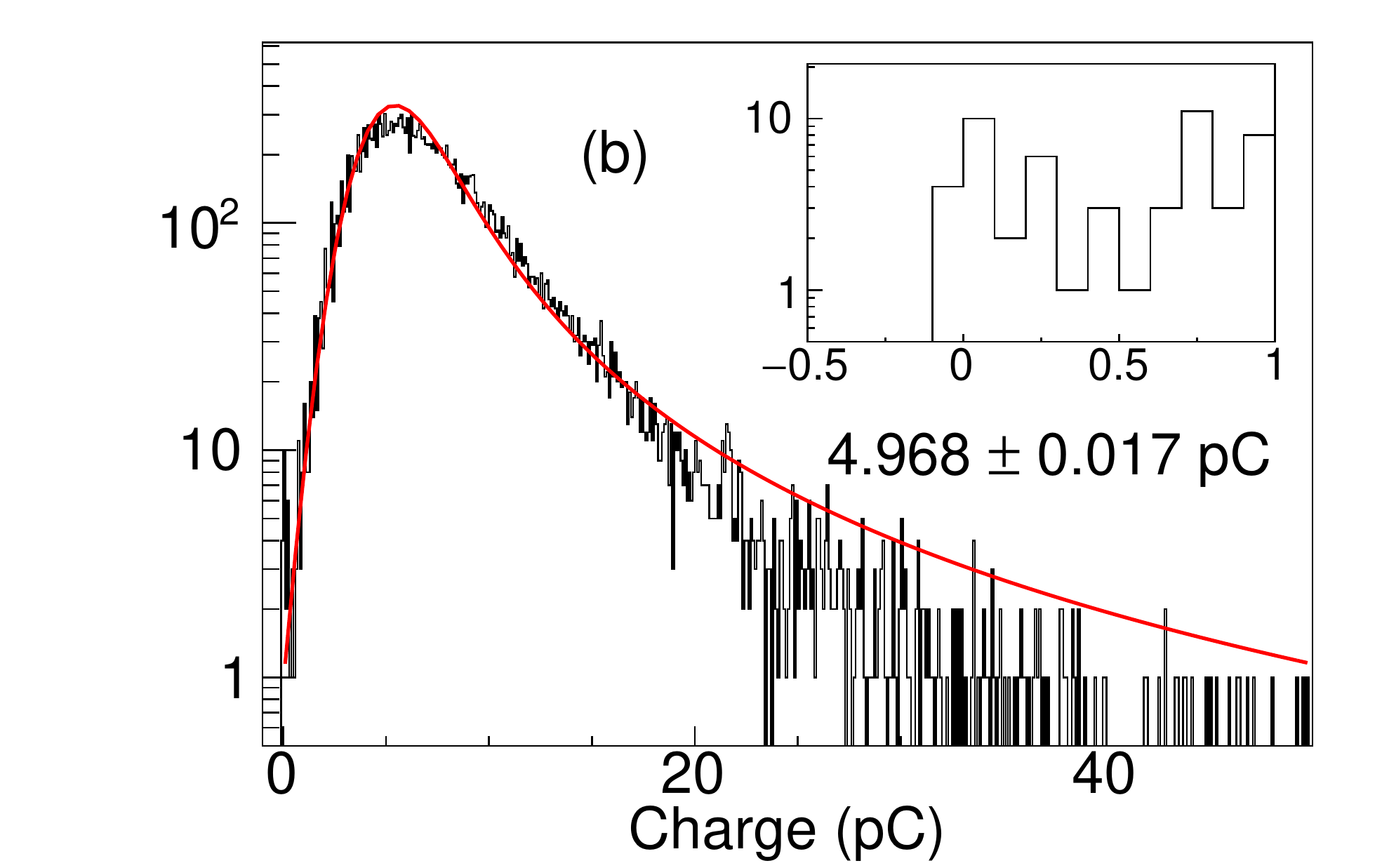}
\caption{\label{fig:cosmicevt} (a) An example of a cosmic muon signal from the di-counter setup and (b) Integrated charge distribution for one of the SiPM channels from the setup at $V_{OV}$=2.5\,V, fitted with Gaussian convoluted Landau function. Here 1\,p.e. corresponds to 0.28\,pC.}
\end{figure}

Figure~\ref{fig:cosmicevt}a shows an example of a cosmic muon signal from an SiPM. The full time profile of the signal is recorded for 1024\,ns. The signal is integrated within a 100\,ns time window and pedestal subtraction is done to remove the effect of baseline fluctuations. Figure~\ref{fig:cosmicevt}b shows the integrated charge distribution for one of the SiPM channels from the setup. The distribution was fitted with a Gaussian convoluted Landau function and the most probable value of charge is estimated from the fit. The tail part of the distribution does not fit properly due to the saturation effect of scintillator – a high ionisation density along the track of the particle leads to quenching from damaged molecules and a reduction in the number of the scintillation photons~\cite{birks}.\\

As mentioned above, to calculate the cosmic muon detection efficiency, the SiPM channels A1 and A3 (shown in figure~\ref{fig:dc_sch}) were added together and considered to be a single channel as there will be muon signal either in A1 or A3 in coincidence with the trigger signal as per the criteria of the TYPE1 geometry. Similarly, (A2, A4), (B1, B3) and (B2, B4) were each considered to be a single channel.

\begin{figure}[h]
\centering 
\includegraphics[width=0.98\textwidth]{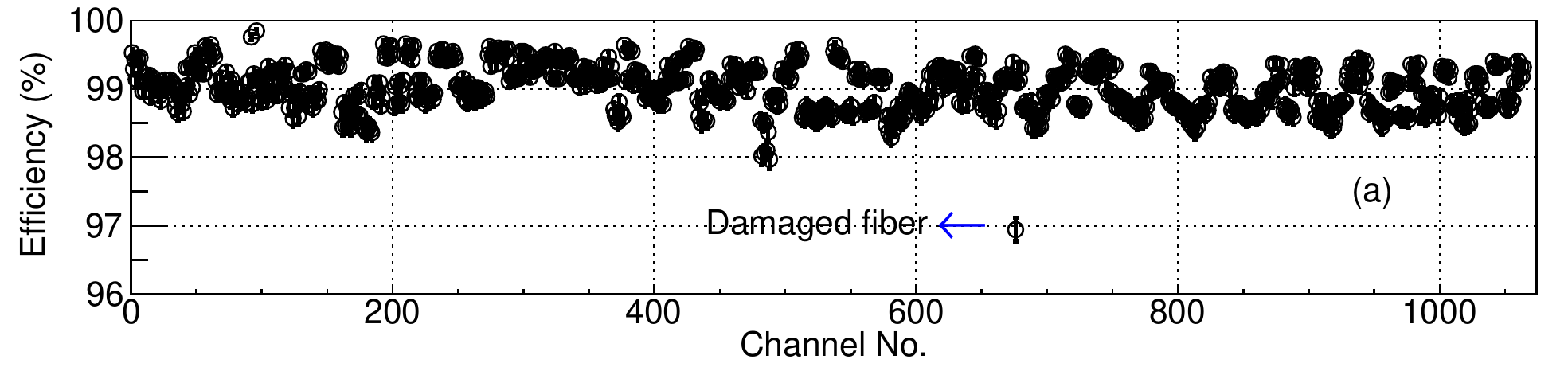}
\includegraphics[width=0.98\textwidth]{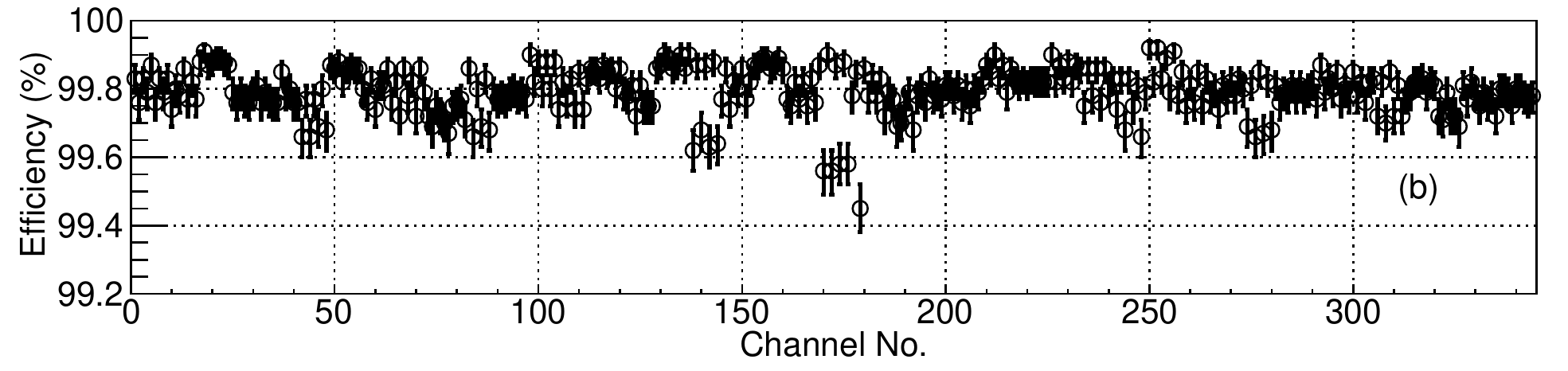}
\caption{\label{fig:effi_sipm} The cosmic muon efficiency measurements at 2.5\,p.e. threshold using (a) TYPE1 geometry for 10\,mm di-counters and (b) TYPE2 geometry for 20\,mm di-counters.}
\end{figure}

The cosmic muon efficiency was estimated at different charge thresholds ($q_{th}$) by calculating the fraction of events crossing that particular $q_{th}$. The cosmic muon efficiency measurements were done at different values of $q_{th}$ for an optimized\footnote{the over voltage at which the required muon detection efficiency is achieved and the noise rate is tolerable.} overvoltage (ov) $V_{ov}$ = 2.5\,V~\cite{mamta1} for all the data sets. Figures~\ref{fig:effi_sipm}(a) and \ref{fig:effi_sipm}(b) show the cosmic muon efficiency measurements at 2.5\,p.e. threshold – the optimized p.e. threshold~\cite{mamta1} – for all the 10\,mm di-counters using the TYPE1 geometry and 20\,mm di-counters using the TYPE2 geometry respectively. There were a few exceptions where the efficiency was extremely low and the charge collection was much less than the average. For instance, in one of the cases, it was observed that the fibre had been damaged and was cracked at the edge of the di-counter. One of the damaged fibre observations is marked in figure~\ref{fig:effi_sipm}(a). The damaged di-counters were repaired by removing the fibre and reusing the di-counter for a shorter length.

\begin{figure}[h]
\centering 
\includegraphics[width=0.98\textwidth]{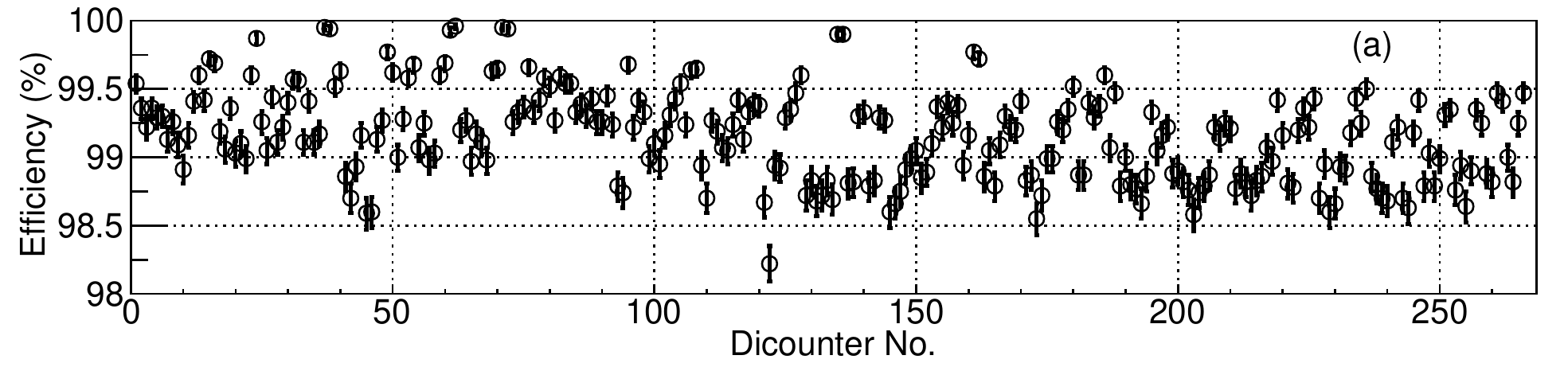}
\includegraphics[width=0.98\textwidth]{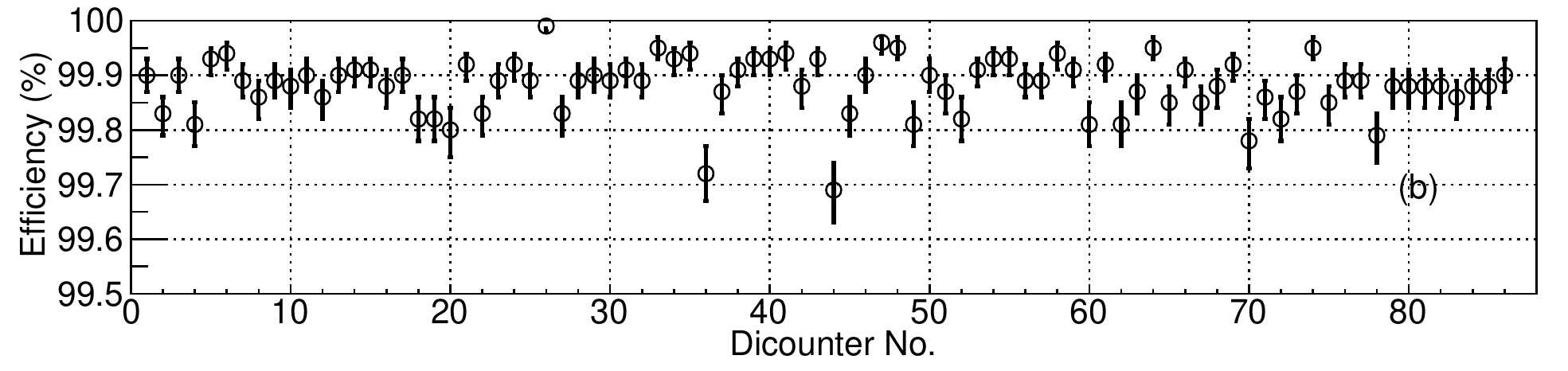}
\caption{\label{fig:sipm_any2} The cosmic muon efficiency measurements for a scintillator when any two out of four SiPMs have signals above 2.5\,p.e. threshold for (a) 10\,mm di-counters using TRIGGER1 method with the TYPE1 geometry and (b)  20\,mm di-counters using TRIGGER1 method with the TYPE2 geometry.}
\end{figure}

The CMVD requires to have a veto efficiency better than 99.99\%. The efficiency will be more than 99.99\% only if the scintillator efficiency – when any two out of four SiPMs in a scintillator have signals above threshold – is at least 99\%. This has been arrived at, after considering the gap between the two adjacent EPS strips and the efficiency of the individual strips~\cite{mamta1}. Figures~\ref{fig:sipm_any2}(a) and \ref{fig:sipm_any2}(b) show the cosmic muon efficiency measurements for a scintillator when any two out of four SiPMs have signals above 2.5\,p.e. threshold for all the 10\,mm di-counters (using TYPE1 geometry, TRIGGER1 method) and 20\,mm di-counters (using TYPE2 geometry, TRIGGER1 method) respectively. Clearly from figure~\ref{fig:sipm_any2}(a), for the 10\,mm di-counters there were many di-counters which did not satisfy the required efficiency criteria.

\begin{figure}[h]
\centering 
\includegraphics[width=0.35\textwidth]{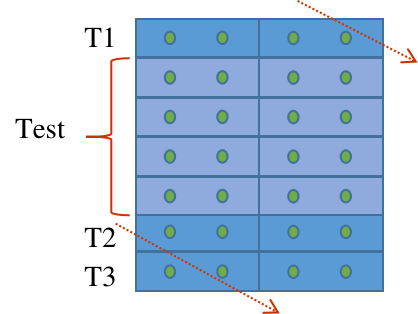}
\caption{\label{fig:falsetrig} Schematic of the side view of the di-counter cosmic muon test setup.}
\end{figure}

The observed cosmic muon detection efficiency in these studies was found to be less than those from the previous studies~\cite{mamta1}. In the TYPE1 geometry, there are chances of FALSE triggers due to multiple muons i.e. it might happen that one muon passes through the top trigger di-counter only and another muon passes through the two bottom trigger di-counters as shown in figure~\ref{fig:falsetrig}. This way, a three-fold cosmic muon coincidence trigger is generated but there is no muon signal in the di-counters under test. In the case of the TYPE2 geometry, the false triggers will be reduced as compared to the TYPE1 geometry and hence the 20\,mm di-counters show better efficiency as compared to the 10\,mm di-counters. To confirm the explanation of false triggers for inefficiency, the data was taken for a few di-counters with the TYPE3 geometry and a different trigger criteria.

\begin{figure}[h]
\centering 
\includegraphics[width=0.4\textwidth]{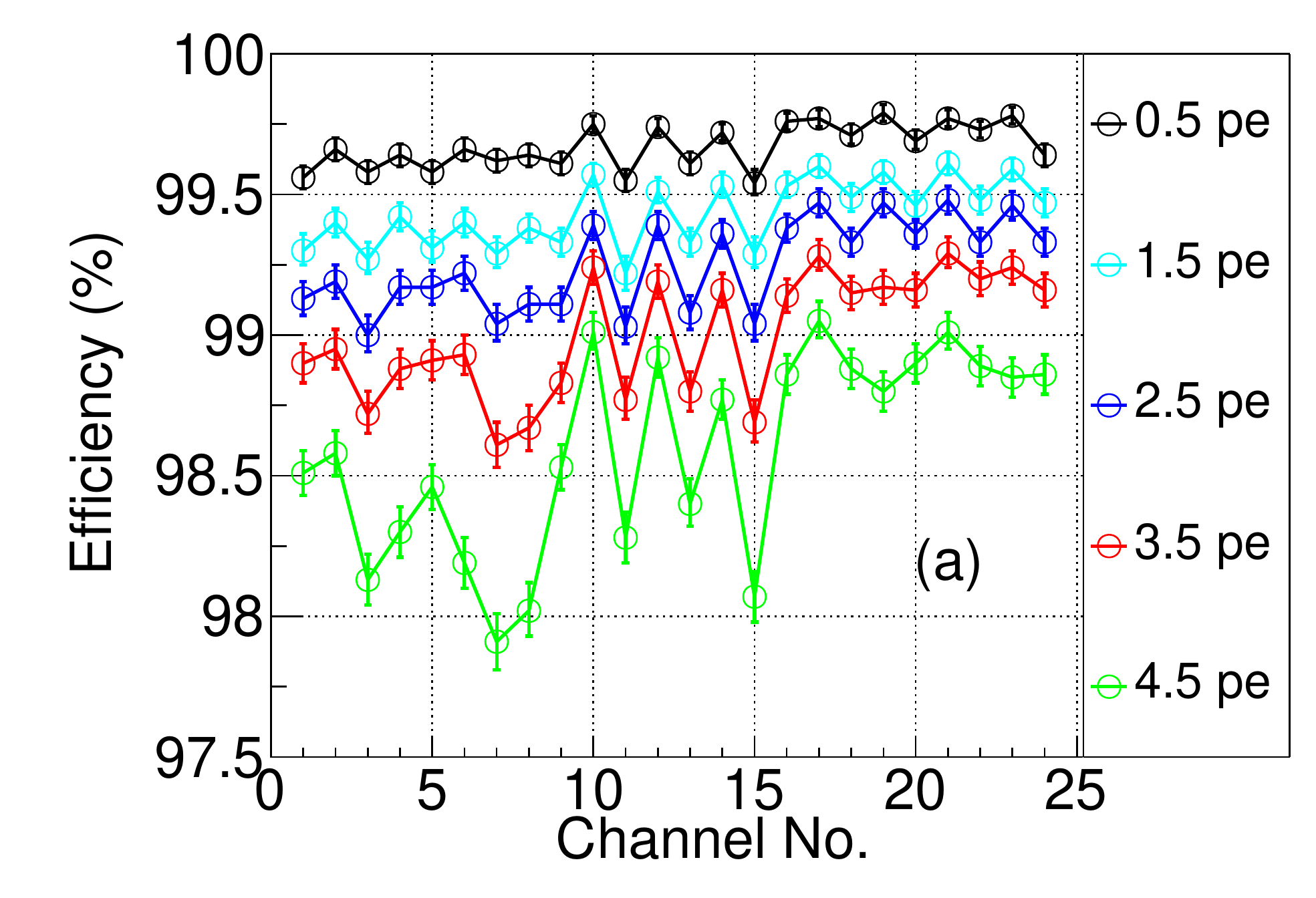}
\hspace{0.02\textwidth}
\includegraphics[width=0.4\textwidth]{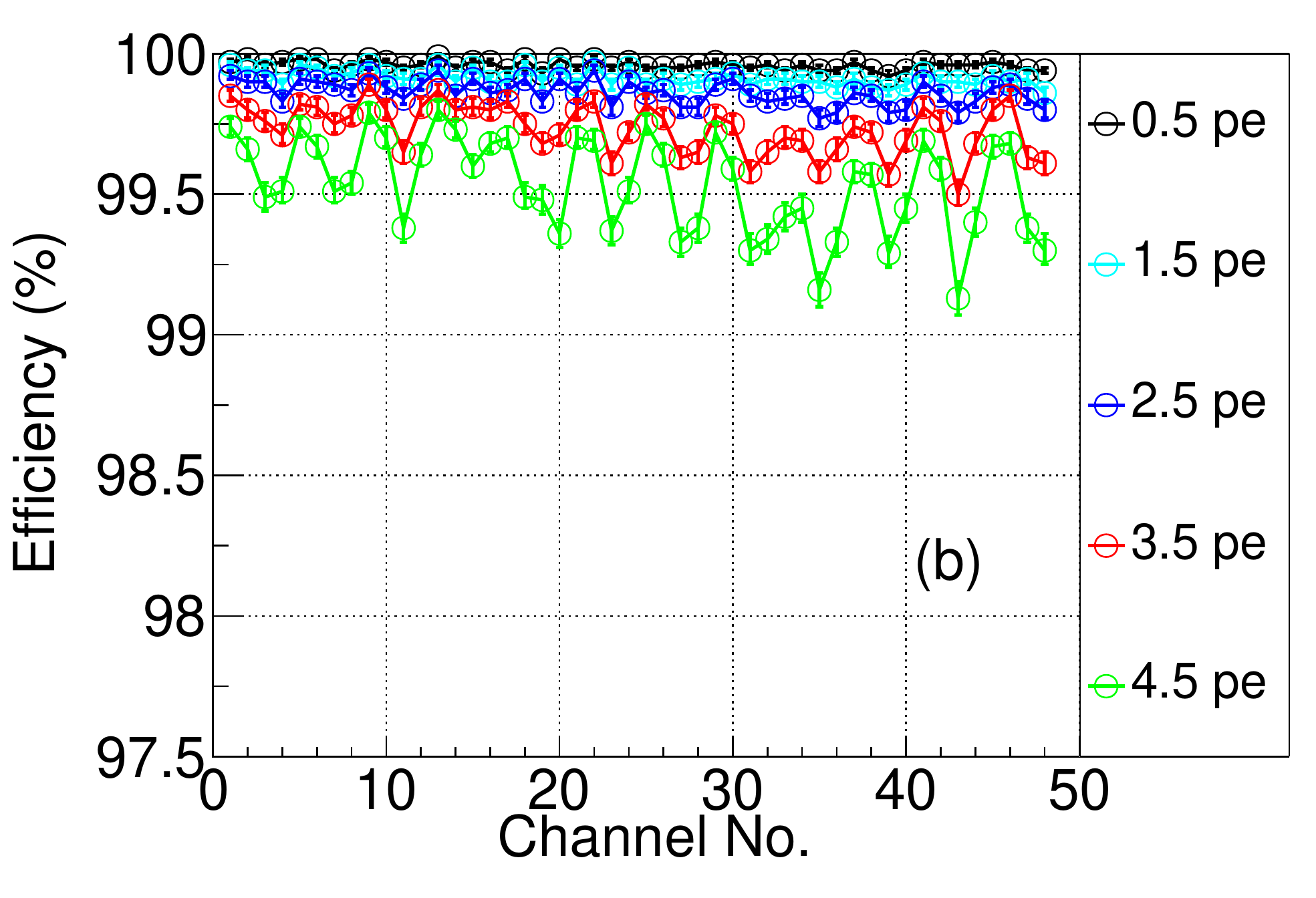}
\caption{\label{fig:effi_sipm_T2_geom}The cosmic muon efficiency measurements from 10\,mm di-counters using (a) TYPE1 geometry with TRIGGER1 method and (b) TYPE3 geometry with TRIGGER2 method, for different p.e. thresholds.}
\end{figure}

The trigger criteria as mentioned in section~\ref{sec:trigger} (let's call it TRIGGER1) was used so that both the EPS strips in the di-counter can be tested simultaneously in a single run. However, testing one EPS strip of a di-counter at a time would have been a better way. So, for a few di-counters, only one EPS strip of a di-counter was tested at a time i.e. the trigger was generated when both the SiPMs of one of the EPS strips, for all the three trigger di-counters, had a muon signal (let's call it TRIGGER2). For these few di-counters, the TRIGGER2 method was studied with the TYPE3 geometry. Figure~\ref{fig:effi_sipm_T2_geom} shows the efficiency measurements from both (TYPE1 geometry, TRIGGER1 method) and the TYPE3 geometry using the TRIGGER2 method for a total of six di-counters. From figure~\ref{fig:effi_sipm_T2_geom}, it is clear that there is an improvement in the efficiency for the TYPE3 geometry using the TRIGGER2 method.

Out of the two trigger methods and three geometries, the efficiency is the best in the case of the TYPE3 geometry and TRIGGER2 method. The correction factor for the efficiency measurements was estimated using the ratio of cosmic muon efficiency measurements from the TYPE3 geometry with TRIGGER2 data sets to that from the TYPE1 geometry with TRIGGER1 data sets for the 10\,mm di-counters. The average efficiency correction factor at 2.5\,p.e. threshold was found to be 1.012 for individual SiPM channels and 1.008 for a scintillator when two out of four SiPMs in a scintillator have signals.

The efficiency of the 20\,mm di-counters when at least two out of four SiPMs in a scintillator have signals above 2.5\,p.e. threshold, is shown in figure~\ref{fig:sipm_any2}(b) and it is more than 99\% for all the scintillators even without adding the correction factors similar to those used for the 10mm DCs. For the 10\,mm di-counters, the scinitllator efficiency is less than 99\% for many di-counters as shown in figure~\ref{fig:sipm_any2}(a), but after compensating for false trigger using the calculated average efficiency correction factor, the 10\,mm di-counters also satisfy the efficiency requirement. The minimum efficiency for any 10\,mm di-counter was observed to be 98.22\% and after implementing the correction factor, it improves up to 99.01\% which is above the minimum requirement i.e. 99\%. Thus, both 10\,mm as well as 20\,mm di-counters satisfy the efficiency requirement of the CMVD. 

\section{Light tightness test for tiles}
Though the tiles are wrapped with a light proof material, it is required to measure for any light leakage in the detectors. To test this, a tile was mounted with four CMBs on one side, while the other side was covered with a black cloth and a layer of Tedlar paper in order to prevent light from entering through the WLS fibres from the open end. The SiPMs used for tile leak test were tested separately beforehand and the SiPM waveforms were recorded. This was done by operating the SiPMs standalone i.e. without scintillators, in a dark environment, and using a random trigger.

Using the data obtained, the thermal noise rates of these SiPMs were estimated at the optimised $V_{ov}$ and at different charge thresholds ($q_{th}$). These SiPMs were then mounted on the tiles using the SMBs. A random trigger was generated and the SiPM waveforms were again recorded at the optimised $V_{ov}$. The noise rates were estimated using this data set also and a comparison was made with the thermal noise rate measurements from the standalone SiPM setup. If there was a light leak, there would have been a significant increase in the noise rate due to the external light as compared to the standalone SiPM setup. Also, a bright flashlight was shone over the entire surface and on the edges of the tile to check if there is any increase in the bias current, none was seen except when the black sheet was torn in a few cases, which were repaired using Tedlar paper and black tape.\\

\begin{figure}[h]
\centering 
\includegraphics[width=0.98\textwidth]{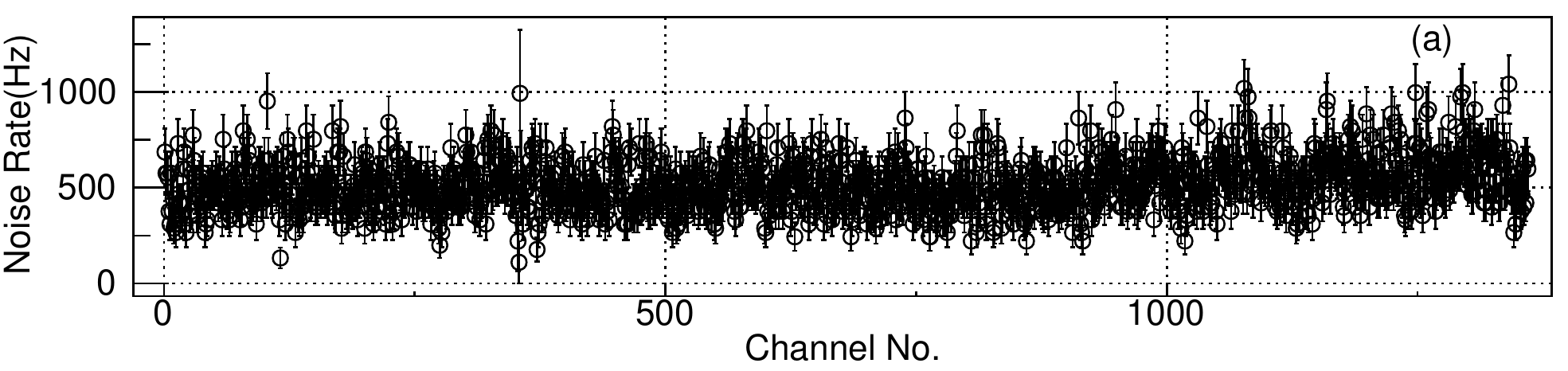}
\includegraphics[width=0.98\textwidth]{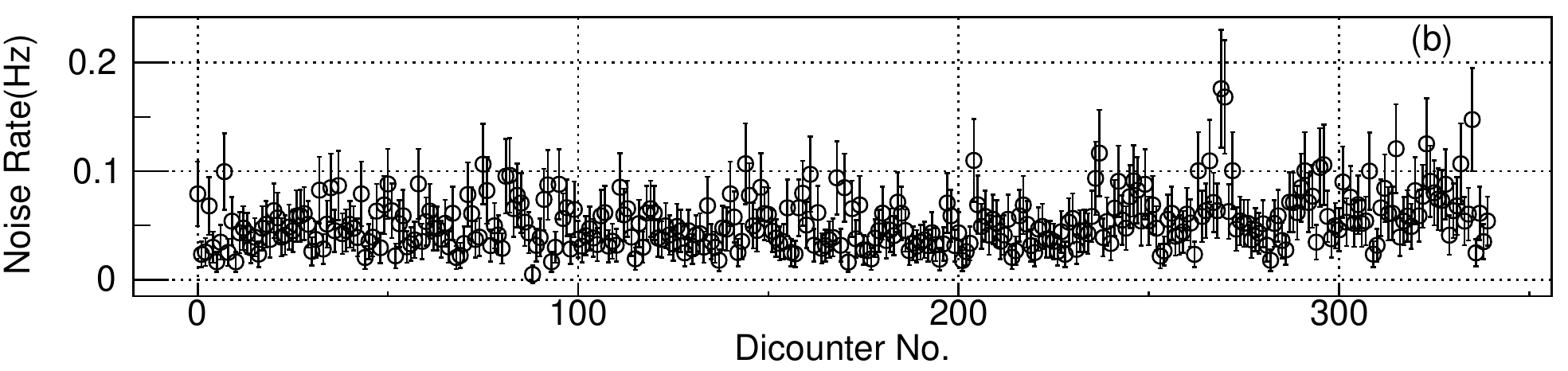}
\caption{\label{fig:tilenoise} The noise rate measurements at 2.5\,p.e. threshold (a) for individual SiPM channel when mounted on different tiles and (b) when any two out of four SiPMs have noise signal simultaneously.}
\end{figure}

Figure~\ref{fig:tilenoise}(a) shows the SiPM noise rates of individual SiPM channels at 2.5\,p.e. threshold. Figure~\ref{fig:tilenoise}(b) shows the coincidence noise rates at 2.5\,p.e. threshold i.e. when any two out of four SiPMs in a scintillator have a noise signal simultaneously. Both these rates were estimated for the SiPMs and corresponding scintillators for different tiles. The noise rates of the individual SiPMs and the coincidence noise rates are within the tolerable noise level~\cite{mamta1} for the CMVD.

\section{Conclusion}
The main aim of this study was to test all the EPS strips to be used for the Cosmic Muon Veto Detector (CMVD), which is being designed by the INO collaboration to cover the mini-ICAL detector which is operational at the IICHEP transit campus, in Madurai in South India. The cosmic muon detection efficiency measurements for the 10\,mm and the 20\,mm di-counters show that the minimum scintillator efficiency requirement of at least 99\% is satisfied by all the di-counters to achieve the main specification of the CMVD to have 99.99\% veto efficiency. All these tested di-counters were used to produce the tiles required for the CMVD. All the tiles have been tested for light tightness and any light leaks were repaired using Tedlar paper and black tape. The tiles are ready for installation and commissioning of the CMVD around the mini-ICAL.

\acknowledgments
We sincerely thank Piyush Verma, Darshana Gonji, Santosh Chavan, Vishal Asgolkar, E. Yuvaraj, Jayakumar Ponraj, Karthikk K.S., Umesh L. and  Jim M John for their support and help during the work. We would also like to thank all other members of the INO collaboration for their valuable inputs. We thank the TIFR central workshop team for helping us in prototyping and manufacturing some of the di-counter components. A special thanks to Dr. Craig Group and Eric A. Fernandez from University of Virginia, USA, for sharing the know-how of the EPS di-counter assembly. We also thank Fermilab for providing us the extruded plastic scintillators.



\end{document}